\documentclass[journal, twocolumn]{IEEEtran}
\usepackage{cite}
\usepackage{amsmath,amssymb,amsfonts,bm}
\usepackage{dsfont}
\usepackage{algorithmic}
\usepackage{graphicx}
\usepackage{textcomp}
\usepackage{xcolor}
\usepackage{color}
\usepackage{tabularx}
\def\BibTeX{{\rm B\kern-.05em{\sc i\kern-.025em b}\kern-.08em
    T\kern-.1667em\lower.7ex\hbox{E}\kern-.125emX}}

\graphicspath{{./fig/}}

\newcommand{\qe}{{\bf e}}

\newcommand{\qh}{{\bf h}}

\newcommand{\qp}{{\bf p}}
\newcommand{\qq}{{\bf q}}

\newcommand{\qu}{{\bf u}}

\newcommand{\qw}{{\bf w}}

\newcommand{\qz}{{\bf z}}

\newcommand{\qH}{{\bf H}}
\newcommand{\qI}{{\bf I}}

\newcommand{\be}{\begin{equation}} \newcommand{\ee}{\end{equation}}
\newcommand{\bea}{\begin{eqnarray}} \newcommand{\eea}{\end{eqnarray}}

\begin{document}

\title{Deep Learning Based Predictive Beamforming Design}
\author{Juping Zhang,
        Gan Zheng,~\IEEEmembership{Fellow,~IEEE,} Yangyishi Zhang,
         Ioannis Krikidis,~\IEEEmembership{Fellow,~IEEE,}
       and Kai-Kit Wong,~\IEEEmembership{Fellow,~IEEE}
 \thanks{This work was supported in part by the CELTIC-Next Project C2019/2-5 Artificial Intelligence-enabled Massive Multiple-input multiple-output (AIMM), the European Regional Development Fund and the Republic of Cyprus through the Research and Innovation Foundation, under the project INFRASTRUCTURES/1216/0017 (IRIDA), and the UK Engineering and Physical Sciences Research Council (EPSRC) grant EP/T015985/1.}
  \thanks{J. Zhang and K.-K. Wong are with the Department of Electronic and Electrical Engineering, University College London, London, WC1E 6BT, UK (Email: \{juping.zhang,kai-kit.wong\}@ucl.ac.uk).}
     \thanks{G. Zheng is with the School of Engineering, University of Warwick, Coventry, CV4 7AL, UK (Email: gan.zheng@warwick.ac.uk).}
     \thanks{Y. Zhang is with the    British Telecom, Adastral Park, Martlesham Heath, Ipswich, IP5 3RE. (E-mail: yangyishi.zhang@bt.com).}
  \thanks{I. Krikidis is with the    Department of Electrical and Computer Engineering,  University of Cyprus, 1678 Nicosia Cyprus. (E-mail: krikidis@ucy.ac.cy).}
 \thanks{For the purpose of open access, the authors have applied for a Creative Commons Attribution (CC BY) licence to any Author Accepted Manuscript version arising.}
}
\maketitle

\begin{abstract}
   This paper investigates deep learning techniques to predict transmit beamforming based on only historical channel data without current channel information in the multiuser multiple-input-single-output downlink. This will significantly reduce the channel estimation overhead and improve the spectrum efficiency especially in high-mobility vehicular communications. Specifically, we propose a joint learning framework that incorporates channel prediction and power optimization, and produces prediction for transmit beamforming directly.  In addition, we propose to use the attention mechanism  in the Long Short-Term Memory Recurrent Neural Networks to improve the accuracy of channel prediction. Simulation results using both a simple  autoregressive  process model and the more realistic 3GPP spatial channel model verify that our proposed predictive beamforming scheme can significantly improve the effective spectrum efficiency compared to traditional channel estimation and the method that separately predicts channel and then optimizes beamforming.
 \end{abstract}

\begin{IEEEkeywords}
channel prediction, beamforming, deep learning.
\end{IEEEkeywords}

\section{Introduction}
Timely and accurate channel state information (CSI)  is essential to exploit the full potential of  multiuser multi-antenna systems by designing the optimal transmission strategies such as beamforming, but it is challenging to obtain in practice. Traditionally downlink CSI is obtained at a base station (BS) via either feedback from users, or channel estimation via uplink pilots by using the channel reciprocity. Both methods introduce significant overhead, extra error and latency, and as a result the CSI at the BS becomes outdated for beamforming design especially in high-mobility scenarios, e.g., unmanned aerial vehicle and vehicle-to-everything communications.

A more efficient channel acquisition method is to predict channels based on historical CSI data by exploiting the temporal correlation. There has been a large body of research on  channel prediction. Early works assume the accurate channel model such as the autoregressive (AR) process   or the long-term channel statistics is available, and typically  Kalman filtering \cite{Long-range} is employed to estimate the AR coefficients. However, practical channels may not be characterized by analytical models and they could be non-stationary, which degrade the performance of model-based prediction methods.  

 Recently deep learning based channel  prediction has received much attention for its ability to learn CSI   from data without prior knowledge about channel models. Starting from a single-antenna system, an efficient long-short term memory (LSTM) network, a type of recurrent neural
network (RNN), is proposed in \cite{Parameter-Free} for CSI prediction and adaptation in  non-stationary changing channels. Going further,  a data-driven receiver architecture that   reduces the pilot overhead is designed in \cite{tvt_prediction}, following RNN-based channel prediction.
The work in \cite{access_prediction} begins by designing and conducting a measurement campaign to collect IQ  samples  of the IEEE 802.11p transmission and extract CSI in various real-world vehicular environments. The LSTM method is then employed to predict future CSI and received signal levels which is verified by trace-based evaluation.
  Deep learning based channel prediction is extended in \cite{ML-aging} to massive multiple-input-multiple-output (MIMO) systems, which improves the channel prediction quality for both low and high mobility scenarios.
    A comparative study on a vector Kalman filter (VKF) predictor and a deep learning based   predictor for massive MIMO systems using   the spatial channel model (SCM) is carried out in \cite{massive-KF-ML}, and it is shown that both can achieve substantial gain over the outdated channel in  terms of the channel prediction accuracy and data rate.
   Following channel prediction, the beamforming optimization has also been studied. A deep learning approach is adopted in \cite{predict_UAV}  to predict the angles between the unmanned aerial vehicle (UAV) and the user equipment in the presence of jittering due to the inherent random wind gusts, such that the UAV and the UE can prepare the transmit and receive beams in advance.
A versatile unsupervised deep learning based predictive beamforming design is proposed in \cite{predict_ISC} in vehicular networks, which implicitly learns the features of historical channels and directly predicts the beamforming matrix to be adopted for the next time slot. The proposed method not only guarantees the required sensing performance, but also achieves a satisfactory sum-rate.
   Predictive beamforming is studied for dual-functional radar-communication (DFRC) systems in vehicular networks \cite{prediction_radar}, in which a novel message passing algorithm based on factor graph is proposed to estimate the motion parameters of vehicles, and the beamformers are then designed based on the predicted angles for establishing the communication links.

Different from the aforementioned works that focus on channel prediction only or have assistance from radar,   we   propose a deep learning framework that takes historical CSI data as input and directly predicts the beamforming solution for future channels to maximize the sum rate performance of a multiuser multi-antenna system.  The framework incorporates an LSTM-based channel prediction module and a power optimization module which helps reconstruct the beamforming vectors using hybrid supervised-unsupervised learning. Furthermore, we propose to use the attention mechanism in the LSTM network such that the impact of historical channels in different coherent intervals will be correctly reflected in the channel prediction and this thus improves the performance of beamforming prediction.

The remainder of this paper is organized as follows. Section II introduces the system model and  the problem formulation.
Section III presents the proposed deep learning framework for predictive beamforming. Simulation results are given to validate the proposed method in Section IV and we conclude our work in Section V.

{\underline{\it Notation}:} All boldface letters indicate vectors (lower case) or matrices (upper case). The superscripts   $(\cdot)^\dag$ and  $(\cdot)^{-1}$ denote  the  conjugate transpose and the matrix inverse, respectively. $\mbox{vec}(\cdot)$ denotes the vector operation of a matrix. The identity matrix is denoted by $\mathbf{I}$. $\|\mathbf{z}\|$  denotes  the $L_2$   norms of a complex vector $\qz$.

\section{System Model and Problem Formulation}
\subsection{System Model}
We consider a  multi-input   single-output (MISO)   downlink system  in which a BS with  $N_t$-antennas serves $K$ single-antenna users that employ single-user detection.
Suppose $s_k$ is the transmit signal to the user $k$ with unit power  and the BS transmits with a total power $P_T$.
 The transmitted data symbol $s_k$   is mapped onto the antenna array elements by multiplying the beamforming vector $\mathbf{w}_k \in \mathbb{C}^{N_t\times 1}$. The received signal at the user $k$ can be expressed as
\begin{align}\label{sys1}
y_k= \mathbf{h}_{k}^\dag \mathbf{w}_k s_k  +  \sum_{j\neq k}\mathbf{h}_k^\dag \mathbf{w}_js_j  +n_k,
\end{align}
where $\mathbf{h}_{k} \in \mathbb{C}^{N_t\times 1}$ is the channel between the BS and the user $k$, the second term represents the interference and $n_k$ denotes the additive white Gaussian noise (AWGN) component with zero mean and variance $\sigma^2_k$.
Therefore, the signal-to-interference-plus-noise ratio (SINR) that measures quality of the data detection   at the $k$-th user is given by
\begin{equation}
\Gamma_k= \frac{  |\mathbf{h}_k^\dag \mathbf{w}_k|^2}{\sum_{j\neq k} |\mathbf{h}_{k}^\dag\mathbf{w}_j|^2+\sigma^2_k}.
\label{eq:Gamma}
\end{equation}
We choose the sum rate as the system performance metric to maximize and the resulting problem is expressed as
\be\label{eq:P0}
 \max_{\{\qw_k\}}R \triangleq \sum_{k=1}^K \log(1+ \Gamma_k), \mbox{s.t.} \sum_{k=1}^K \|\qw_k\|^2\le P_T.
\ee
{{ The sum rate optimization problem in \eqref{eq:P0} is nonconvex  and  the standard approach to find its suboptimal solution is to use  the weighted minimum mean squared error (WMMSE) algorithm \cite{christensen2008weighted} assuming the CSI $\{\qh_k\}$ is available which normally relies on pilot-based channel estimation and introduces substantial overhead.}}

\subsection{Problem Formulation}
 In this paper, we adopt a hybrid channel-estimation-prediction scheme  to solve the problem \eqref{eq:P0}, in which we have CSI estimation of  a certain number of channels and then use it to predict CSI of some future channels. Specifically, we assume the time horizon is divided into frames of $N+P$ coherent intervals, and within each frame, the CSI of the first $N$ coherent intervals is available, and the CSI of  the rest $P$ coherent intervals will be predicted without estimation.

 The known CSI estimation is written as
 \be\label{eq:CSI:est}
    \hat\qh_k (n) =\qh_k(n) + \qe_k(n), \forall n=1, \cdots, N,
 \ee
 where $\qh_k(n)$ is the true CSI of user $k$ at the $n$-th coherent interval and $\qe_k(n)$ is the channel estimation error that follows the complex Gaussian distribution with zero mean and variance matrix of $\sigma^2_{e,k} \qI$. We assume the least-square channel estimation is used, so $\sigma^2_{e,k}$ depends on pilot transmission power.

 The CSI prediction is expressed as
 \be
    \tilde\qh_k(N+m) = \mathds{M}(\{\hat\qh_k (n)\}, \forall n=1, \cdots, N), \forall m=1, \cdots, P,
 \ee
 with $\mathds{M}(\cdot)$ being the mapping from the known channel estimation to the channel prediction.

 With the hybrid scheme, our aim is to solve the problem \eqref{eq:P0} by predicting the beamforming solutions directly for those  $P$ unkown channels given the $N$ known channels.

\section{Predictive Beamforming Solution}
\subsection{The General framework}
 Our proposed deep learning based framework to predict the beamforming solution is illustrated in Fig. \ref{fig:NN} below.
\begin{figure}[h]
  \centering
  \includegraphics[scale=0.48]{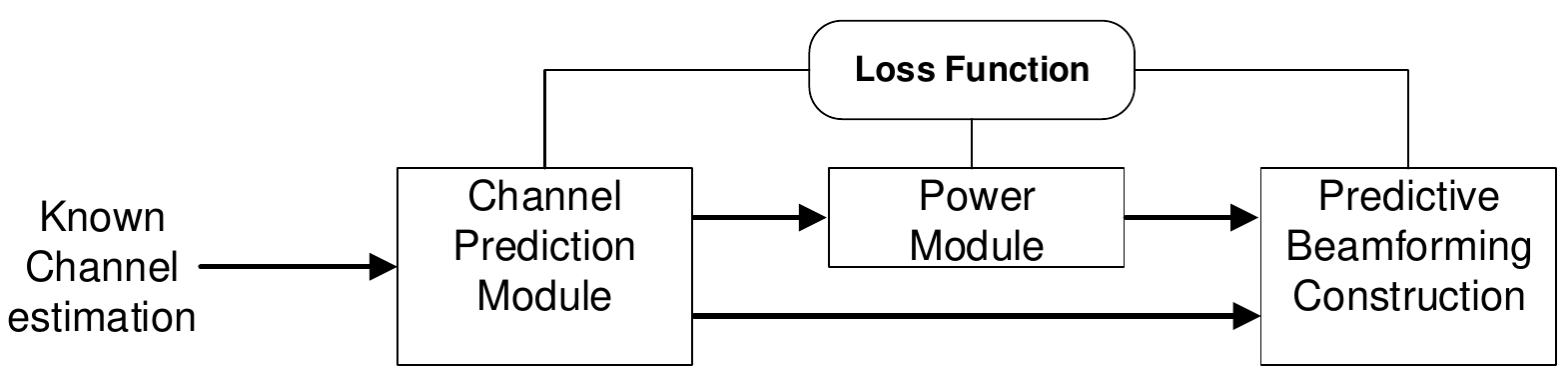}
  \caption{The proposed framework for predictive beamforming.}\label{fig:NN}
\end{figure}

 The proposed framework includes three key modules. The first module is an neural network (NN) for channel prediction, which realizes the mapping $\mathds{M}(\cdot)$ from the input   known channel estimations $\hat\qh_k (n), n=1,\cdots,N$ to the predicted channel $\tilde\qh_k(N+m), m=1, \cdots, P$. Supervised learning is adopted and the associated loss function is
\begin{equation}\label{loss1}
 L_H =\frac{1}{2LK}\sum_{l=1}^L\sum_{k=1}^K\sum_{m=1}^P
   {\left(\| \qh_k(N+m)^{(l)}- \tilde\qh_k(N+m)^{(l)}\|^2\right)},
\end{equation}
where $L$ is the number of batches, and the superscript $^{(l)}$ denotes the index of the batch. Details of the channel prediction module will be introduced in the next subsection.

  The second module is an NN to predict power vectors in order to facilitate the prediction of beamforming solution. With the predicted channel, the beamforming solution could be inferred directly using supervised learning, but the high dimensional beamforming will make the training challenging and reduce the accuracy of inference. Instead, we exploit the following parameterized structure of the beamforming solution for the sum rate maximization problem:
 \be\label{eq:w}
     \tilde\qw_k = \sqrt{p_k}\frac{\left(\qI + \sum_{j=1}^K \frac{q_j}{\sigma_j^2}  \tilde\qh_{j}\tilde\qh_{j}^\dag   \right)^{-1}\tilde\qh_k}{\| \left(\qI + \sum_{j=1}^K\frac{q_j}{\sigma_j^2}  \tilde\qh_{j}\tilde\qh_{j}^\dag   \right)^{-1}\tilde\qh_k\|},
\ee
where $\{\qq\}$ and  $\{\qp\}$ are the uplink and downlink power vectors, respectively, with the same total sum power of $P_T$. This structure is adapted from the result for perfect CSI \cite{bjornson2014optimal}. It can be seen that once the channel prediction is obtained, we can employ an NN to predict the power vectors $\{\qq\}$ and  $\{\qp\}$ using supervised learning, and then reconstruct the beamforming solution using \eqref{eq:w}. The labelled data can be obtained using the WMMSE algorithm \cite{christensen2008weighted}. For simplicity, we use  fully connected layers for the power module with details provided in Section IV. Suppose the associated loss is given by
\bea\label{loss_power}
  L_P &=&\frac{1}{2LK}\sum_{l=1}^L\sum_{m=1}^{P}{\left(\|\qq^{(N+m,l)}-\tilde{\qq}^{(N+m,l)}\|^2\right)}\\
   &+& {\left(\|\qp^{(N+m,l)}-\tilde{\qp}^{(N+m,l)}\|^2\right)},
\eea
where $\tilde{\qq}$ and $\tilde{\qp}$ are the output of the power NN. 

The next module will leverage the predicted channel and power to construct the predictive beamforming solution using \eqref{eq:w}. To improve the end performance of maximizing the sum rate, we also consider to incorporate the sum rate expression $R(\{\tilde h_k, \tilde\qw_k\})$ in \eqref{eq:P0} into the loss function for training the overall NN.  
The overall loss function is thus given as a weighted sum below:
\be\label{eq:loss}
    L = L_H + L_P - b R(\{\tilde h_k, \tilde\qw_k\}),
\ee
where $b$ is a positive coefficient. {{The novel design of the loss function \eqref{eq:loss} reflects the fact that our proposed solution is a joint learning framework that incorporates channel prediction and power optimization, while it produces prediction for transmit beamforming directly.}} 
The rest of this section is devoted to the development of the channel prediction module.

\subsection{Attention-based LSTM Channel Prediction}
 Channel prediction is to predict  future CSI given historical CSI data by exploiting the temporal correlation between them.  The  RNN is a well-suited machine learning technology to predict time series data \cite{nn-Overview}. 
 However, standard RNNs have the issues of vanishing and exploding gradients during back-propagation, which makes predicting long time series data sequences challenging.  LSTM is one of the most successful variants of   RNNs to predict the correlate time series data and can solve the issues of vanishing and exploding gradients  \cite{nn-Overview}, so it is adopted in this paper to predict the temporal correlated channel.

 An LSTM is composed of a memory cell which can store data for long periods. The flow of information into and out of the cell is managed by three gates. Specifically, the  forget gate  determines what information from the previous state cell will be memorized and what information will be removed that is no longer useful; the  input gate  determines which information should enter the cell state; and the  output gate  determines and controls the outputs.

 Most existing works in channel prediction use a simple LTSM and only its last hidden state is fed into a fully connected layer to produce the predicted channels.
In this paper we propose an improved solution by introducing the attention mechanism to allow the algorithm to put focus on different historical channels. Specifically,  the attention scheme will put higher weights on more recent channels to improve the future channel prediction.  This is intuitive since not all historical CSI data have the same impact on a future channel, and it is mostly influenced by more recent channels.  The advantage of the proposed attention scheme is that  the weights will be optimized by training and do not need to be predetermined manually.

   {In this paper we propose an improved solution by introducing the attention mechanism to allow the algorithm to put focus on different historical channels. Specifically,  the attention scheme will assign higher weights on more recent channels to improve the future channel prediction. This is intuitive since not all historical CSI data have the same impact on a future channel, and it is mostly influenced by more recent channels. }
 The attention mechanism has gained remarkable success in sequence-to-sequence tasks like language translation and handwritten word recognition \cite{attention}, but has not been used for channel prediction. 

\begin{figure}[h]
  \centering
  \includegraphics[scale=0.48]{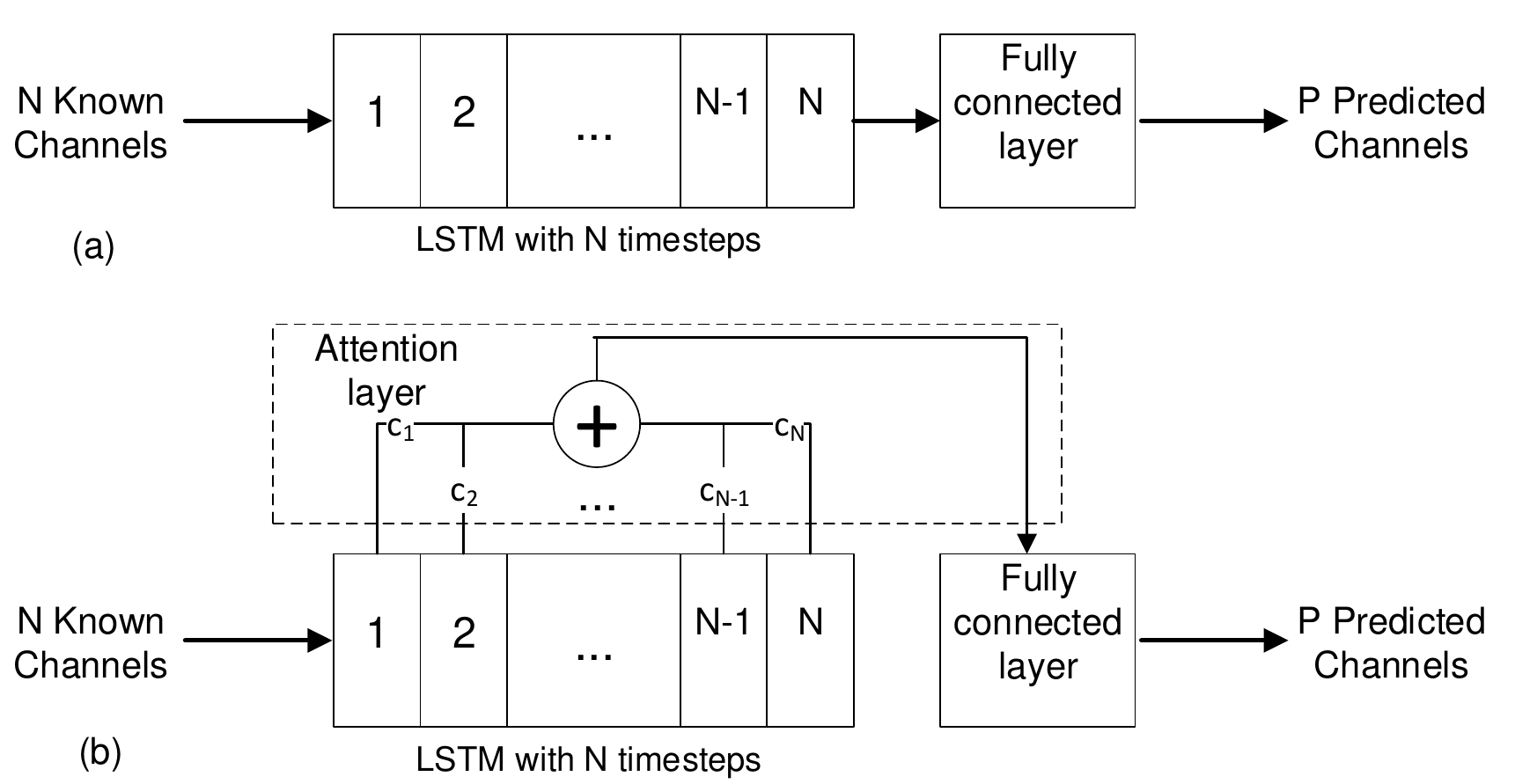}
  \caption{(a) The traditional channel prediction using LSTM without attention; (b) The proposed channel prediction using LSTM with an attention layer.}\label{fig:lstm}
\end{figure}

 The proposed channel prediction module using LSTM with the attention mechanism is depicted in Fig. \ref{fig:lstm}. It has $N$ time steps each corresponding to the known channel estimation in one coherent interval. {{In order to best make  use of the historical data we create an attention layer, which is located between the LSTM and the  fully connected layer.}} This layer assigns a weight $c_n$ to the hidden state output of each time step $n$ with $\sum_{n=1}^N c_n=1$, and then combines the weighted sum of the original hidden states as the LSTM output state. This will allow the importance of channel estimation in different coherent intervals to be correctly characterized. The new output state is then fed into  fully connected layers to produce the predicted channels. Instead of treating the weights $\{c_n\}$ as hyperparameters, one distinct advantage of the proposal scheme is to incorporate them in the overall neural network training, so there is no need to  adjust them manually.  

\subsection{Effective Sum Rate}
 In this subsection we define and analyze the effective sum rate $R_E$ as a performance metric that incorporates the sum rates of estimated channels and predicted channels, which can be written as
\be
    R_E = \frac{(1-\alpha)N}{N+P} R_e + \frac{P}{N+P} R_p,
\ee
where $\alpha$ as the portion of the channel estimation overhead in an coherent interval and therefore it satisfies $\alpha<1$, $R_e$ is the average sum rate for the estimated channels, which can be obtained using the existing neural network method such as that in \cite{dl_bf_zheng} as if the CSI is perfect,  while $R_p$ is the average sum rate for the predicted channels, which is obtained using the proposed predictive beamforming method.  Note that compared to the traditional channel estimation based optimization method, our proposed method reduces the channel estimation overhead by $\alpha P$.  Intuitively, the signalling overhead caused by channel estimation reduces the effective sum rate, so choosing a large $N$ may be beneficial. However, the quality of channel prediction also relies on the number of known channel estimations. Therefore, the effective sum rate is in general not monotonically varying according to $N$ and $P$, and a balance  between low overhead and high-quality prediction needs to be achieved in practice.

\section{Numerical Results}
\subsection{Simulation setup}
In this section, we provide numerical results to validate and evaluate  the performance of the proposed deep learning-based predictive beamforming solution. Unless otherwise specified, we consider a  MISO downlink consisting of $N_t=4$ transmit antennas and $K=3$ users. The transmit signal-to-noise (SNR) $P_T$  and the variance of channel estimation error is $\sigma^2_{e,k}= 1/P_T, \forall k$. The channel estimation overhead is set to be $\alpha=0.1$.  {{Unless otherwise specified, we assume $N=20$ known CSI data are available and $P_T$ is 20 dB.}} In our simulation, we generate 30,000 training labels and 1,000 testing samples, respectively, using the WMMSE algorithm in \cite{christensen2008weighted}.
For the LSTM layer, we use $2N_tKN$ units while three fully connected layers each with $30N_tK$ neurons, relu activation are used for the power NN. $b$ is chosen to be 0.001 in \eqref{eq:loss}.
{{We use Python and Keras in Tensorflow to train the proposed deep learning model. All simulation results are generated by using a computer with an Intel i7-7700 CPU and an NVIDIA Titan Xp GPU.}}
The normalized MSE (NMSE) defined below is used as the performance metric for channel prediction:
\be
    \mbox{NMSE} =  \mathbb{E}\left[\frac{\|\qH - \tilde \qH\|^2}{\|\qH\|^2}\right].
\ee

The following benchmark schemes are considered for comparison:
\begin{itemize}
  \item Predict beamforming without attention. It is the same as the proposed method except that the channel prediction module does not employ the attention mechanism.
  \item Separate optimization, i.e., to use the proposed LSTM-based method with attention to predict the channel, and the zero-forcing (ZF) beamforming is used to optimize the sum rate of both the predicted channels and the known channels. 
  \item Channel estimation followed by ZF beamforming. This is the traditional  estimation scheme with pilot overhead in which no channel prediction is used. It may not always be feasible due to the latency.
  \item Kalman filtering. This scheme uses Kalman filtering for channel prediction and ZF for optimizing the sum rate.
\end{itemize}

\subsection{Channel models}
 We  consider  two different scenarios for modelling the channel dynamics.
\begin{itemize}
 \item   The first scenario is the first-order AR process.   Suppose we collect all users'  channel as $\qH$. In this scenario, the temporal evolution of the channel is given by
   \be\label{eq:KF}
    \mbox{vec}(\qH_{n}) = \beta\mbox{vec}(\qH_{n-1}) + \qu_n,
   \ee
   where $\beta = \mathcal{J}_0(2\pi f_D T_s)$,    $\mathcal{J}_0(\cdot)$ is the zeroth-order Bessel function of the first kind,  $f_D$ is the maximum Doppler frequency shift and  $T_s$ is the sampling duration. $\qu_n$ is the   zero-mean Gaussian excitation noise with covariance matrix $\sigma^2_u\qI$ with $\sigma^2_u = 1-\beta^2$. The composite term $f_D T_s$ denotes the normalized Doppler rate. In the simulation, we choose $f_D T_s=0.005$,  so $\beta = 0.9998$, which corresponds to a slow user velocity of 2.7 km/h  at a frequency of 2 GHz and sampling duration of 1 ms.
    For the Kalman filtering,  an AR model with the prediction order of one in \eqref{eq:KF} is used together with the measure data in \eqref{eq:CSI:est} to estimate the channels $\{\qh_n\}, 1\le n\le N$. While to predict the channels without measurement data, the simple state evolution is used: $\qh_{N+m} = \beta \qh_{N+m-1}, 1\le m\le P$. Note that we have assumed that the  model in \eqref{eq:KF} with the parameter $\beta$ is known when designing the Kalman filtering, while for the proposed prediction method, neither the model nor the parameter is available and it will learn directly from the data.

    \item  The second scenario is  the  3GPP  urban micro SCM  which represents a more realistic but challenging channel environment \cite{scm}. We employ the default SCM parameters in \cite{scm} except that  NumBsElements is $N_t$,  NumMsElements is 1, and NumPaths  is set to 1.
 \end{itemize}

 \subsection{Results}
\begin{figure}[h]
\centering
 \includegraphics[width=2.5in]{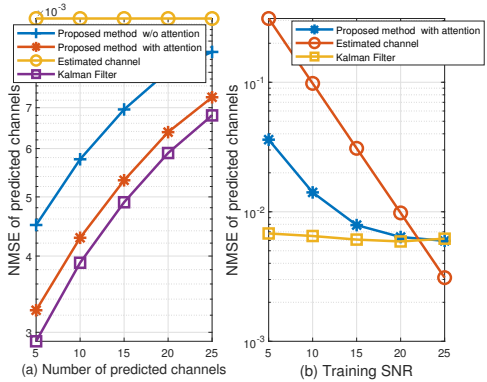} 
  \caption{The NMSE of the predicted channels versus  the number of predicted channels; and (b) SNR, $P=20$.}
\label{fig:NMSE}
\end{figure}

  We first consider the first scenario of AR model.  The NMSE results of the channel prediction  are depicted in Fig. \ref{fig:NMSE}(a) against the number of predicted channels $P$. In general, the more channels to predict, the higher the NMSE is. It is observed that our proposed channel prediction with the attention mechanism is superior to the counterpart without attention, and both achieve much lower NMSE than the traditional channel estimation scheme.  Note that this is because our proposed prediction exploits the temporal correlation of the channel, while the traditional channel estimation does not and it simply uses pilots to estimate the current channel. As the number of predicted channel becomes larger,  the performance of prediction will unavoidably degrade and become  worse than the traditional channel estimation.  In the sequel for our proposed solution we assume   attention   is always used. Kalman filtering achieves the lowest NMSE among all considered schemes by making use of the model information and known parameters,  which are not available to the proposed prediction method.
   The accuracy of the first $N$ channel estimation results will affect the subsequent channel prediction performance, and therefore in Fig. \ref{fig:NMSE}(b) we show the results of NMSE of predicted channels versus the training SNR for the estimated channels when $P=20$. As can be seen, the NMSE of the traditional channel estimation keeps decreasing as the SNR increases; while for the prediction methods, the NMSE saturates when the SNR is above a certain threshold. This is because the performance of channel prediction  is limited by the number of predicted channels no matter how accurate the channel estimation is.

\begin{figure}[h]
\centering
 \includegraphics[width=2.5in]{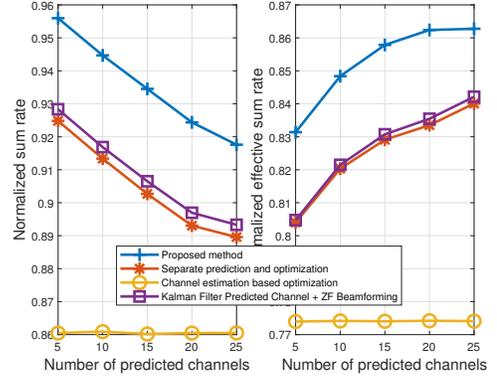}
  \caption{The normalized sum rate results against the number of predicted channels.}
\label{fig:sumrate1}
\end{figure}

 Next, we plot the sum rate and the effective sum rate normalized by the sum rate achieved by the WMMSE solution against the number of predicted channels in Fig. \ref{fig:sumrate1}(a) and Fig. \ref{fig:sumrate1}(b), respectively. It can be seen from Fig. \ref{fig:sumrate1}(a) that our proposed solution achieves the highest sum rate while the sum rate with the traditional channel estimation is the lowest. The separate solution that first predicts the channel and then uses ZF beamforming achieves slightly lower sum rate  than that of Kalman filtering. The same trend can be observed from the Fig. \ref{fig:sumrate1}(b) about the effective sum rate. Because the channel estimation overhead is taken into account, the achievable effective sum rates are lower than those in Fig. \ref{fig:sumrate1}(a). For our proposed solution, the sum rate performance shows ceiling effect as the number of predicted channels increases. As per the analysis in Section III.C, this is because more future channels will reduce the prediction accuracy, and consequently the effective sum rate.

\begin{figure}[h]
\centering
 \includegraphics[width=2.5in]{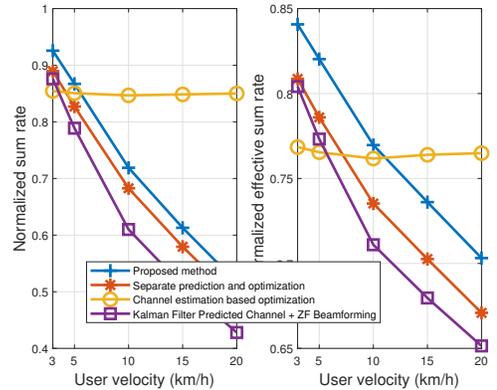}
  \caption{The normalized sum rate results against the user velocity.}
\label{fig:sumrate:speed}
\end{figure}

Next, we examine the effect of the user velocity on the normalized sum rate and the results are provided in Fig. \ref{fig:sumrate:speed}. As expected, the sum rate decreases for all schemes as user mobility increases. Kalman filtering performs well at low velocity, but it cannot track the change of channel dynamics well at high user mobility. Our proposed method achieves much higher effective sum rate than Kalman filter and the separate approach. The channel estimation method clearly outperforms others at high user mobility in theory, but it may not be practical to obtain the channel estimation in time.

\begin{figure}[h]
\centering
 \includegraphics[width=2.5in]{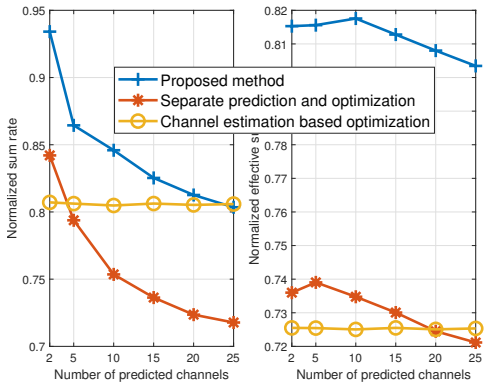}
  \caption{The normalized sum rate results against the number of predicted SCM channels.}
\label{fig:scm}
\end{figure}

 We then consider the second scenario of the urban micro SCM model.   The normalized sum rate and effective sum rate results  are shown in Fig. \ref{fig:scm}.  Kalman filtering is not included in the comparison because in the simulation we found its performance is not satisfactory and this may be because the channel dynamic is too complex for Kalman filtering to predict. We can see from Fig. \ref{fig:scm}(a) that as the number of predicted channels increases, the sum rate of the separate optimization degrades quickly and is even much worse than the traditional channel estimation based optimization. Our proposed solution still achieves the highest sum rate although as the number of predicted channels increases, the performance gap with the traditional solution becomes smaller. This highlights the importance of end to end learning of the predictive beamforming. Fig. \ref{fig:scm}(b) depicts the effective sum rate results. As expected, our proposed solution achieves superior performance, while the performance of the separate solution is worse than the estimation-based optimization when the number of predicted channels is high.  The sum rates of both our proposed solution and the separate solution demonstrate the trend of first increasing and then decreasing. This again validates our analysis in Section III.C that a balance between the reduced overhead and high-quality channel prediction is necessary. For instance, for our proposed solution and the separate solution, the optimal numbers of predicted channels are 10 and 5, respectively.

\begin{figure}[h]
\centering
 \includegraphics[width=2.5in]{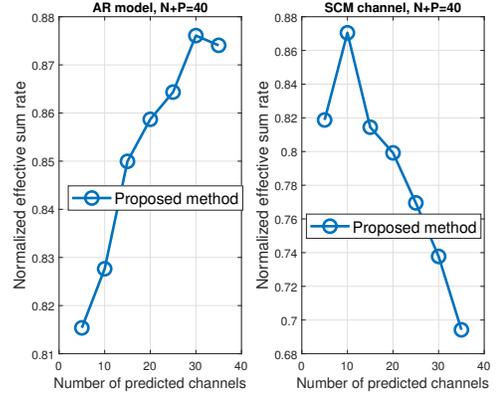}
  \caption{The impact of the number of predicted channels on the normalized sum rate results given a fixed frame length of $N+P=40$.}
\label{fig:tradeoff}
\end{figure}

 Finally, we assess the impact of the number of predicted channels $P$ given a total frame length of $N+P=40$ in Fig. \ref{fig:tradeoff}. There is no quantitative criterion on how to choose the optimal $N$ and $P$. Intuitively, when $N$ is larger, we have more channel information available to predict future channels more accurately, but this also causes higher overhead which will reduce the effective sum rate. A good tradeoff can be obtained   by empirical study for a specific scenario. As can be seen from Fig. \ref{fig:tradeoff}, for the AR-model, the channel is relatively easy to predict, i.e., with a small number  of known channels $N$, the proposed algorithm can predict a large number ($P=30$) of future channels. While for the SCM model, it is more challenging to track the channel evolution, so only a small number ($P=10$) of future channels are predicted in order to achieve a high effective sum rate.

\section{Conclusions}
In this paper, we have studied the predictive beamforming using the deep learning approach in the multiuser MISO downlink. A general framework that predict the beamforming solution to maximize the sum rate with historical channel measurement data was proposed. An LSTM with an attention layer was devised to improve the performance of channel prediction. Simulation results  have shown that the proposed deep learning based solution achieves significantly higher effective sum rate over the traditional channel estimation based optimization and the separate prediction and then optimization scheme.

\end{document}